\documentclass[12pt]{article}
\usepackage{amsmath}
\usepackage{graphicx,psfrag,epsf}
\usepackage{enumerate}
\usepackage{natbib}
\usepackage{url} 

\newcommand{\blind}{0}

\usepackage{amsfonts, amssymb, amsthm, amstext, graphicx, epsfig, mathtools, enumerate,rotating,verbatim,subfigure}
\usepackage{tikz}  

\usepackage{algorithm,algorithmic}
\usepackage{float}
\usepackage{prettyref,soul}
\usepackage{tabulary}

\usepackage[CJKbookmarks=true,
            bookmarksnumbered=true,
			bookmarksopen=true,
			colorlinks=true,
			citecolor=blue,
			linkcolor=blue,
			anchorcolor=red,
			urlcolor=blue]{hyperref}
\usepackage{multirow}
\usepackage{tabularx}
\usepackage{booktabs}
\usepackage{hyperref}
\usepackage{bm}
\usepackage{url}

\newtheorem{proposition}{Proposition}[section]
\newtheorem{thm}{Theorem}

\newtheorem{assumption}{Assumption}
 
\newtheorem{example}{Example}[section]

\newrefformat{eq}{(\ref{#1})}
\newrefformat{chap}{Chapter~\ref{#1}}
\newrefformat{sec}{Section~\ref{#1}}
\newrefformat{algo}{Algorithm~\ref{#1}}
\newrefformat{fig}{Fig.~\ref{#1}}
\newrefformat{tab}{Table~\ref{#1}}
\newrefformat{rmk}{Remark~\ref{#1}}
\newrefformat{clm}{Claim~\ref{#1}}
\newrefformat{def}{Definition~\ref{#1}}
\newrefformat{cor}{Corollary~\ref{#1}}
\newrefformat{lmm}{Lemma~\ref{#1}}
\newrefformat{lemma}{Lemma~\ref{#1}}
\newrefformat{prop}{Proposition~\ref{#1}}
\newrefformat{app}{Appendix~\ref{#1}}
\newrefformat{ex}{Example~\ref{#1}}
\newrefformat{exer}{Exercise~\ref{#1}}
\newrefformat{soln}{Solution~\ref{#1}}
\newrefformat{cond}{Condition~\ref{#1}}

\def\text#1{\mbox{\rm #1}}

\newcommand{\diag}{{\rm diag}}

\newcommand{\ie}{\mbox{\sl i.e.\;}}

\newcommand{\wh}{\widehat}

\def\be{\begin{equation}}
\def\ee{\end{equation}}
\def\T{{ \mathrm{\scriptscriptstyle T} }}

\newcommand{\bX}{{\mathbf X}}
\newcommand{\bY}{{\mathbf Y}}

\newcommand{\ba}{{\mathbf a}}

\newcommand{\bfe}{{\mathbf e}}

\newcommand{\bu}{{\mathbf u}}
\newcommand{\bv}{{\mathbf v}}

\newcommand{\bx}{{\mathbf x}}

\newcommand{\bz}{{\mathbf z}}

\newcommand{\bbeta}  {\boldsymbol{\beta}}
\newcommand{\bfeta}  {\boldsymbol{\eta}}

\newcommand{\bOmega}{\boldsymbol{\Omega}}

\newcommand{\bSigma}{\boldsymbol{\Sigma}}

\newcommand{\bPhi} {\boldsymbol{\Phi}}

\newcommand{\bmu} {\boldsymbol{\mu}}

\newcommand{\bLambda} {\boldsymbol{\Lambda}}

\usepackage{float}
\usepackage{xr}

\graphicspath{{fig/}}

\begin{document}

\def\spacingset#1{\renewcommand{\baselinestretch}%
{#1}\small\normalsize} \spacingset{1}


\if0\blind
{
  \title{\bf Regression Modeling of the Count Relational Data with Exchangeable Dependencies}
  \author{Wenqin Du\thanks{wenqindu@marshall.usc.edu}\hspace{.2cm}\\
    Department of Data Sciences and Operations,\\ Marshall School of Business, University of Southern California \\
    Bailey K. Fosdick\thanks{bailey.fosdick@cuanschutz.edu}\hspace{.2cm}\\
    Department of Biostatistics and Informatics, \\
    Colorado School of Public Health\\
    Wen Zhou\thanks{w.zhou@nyu.edu}\hspace{.2cm}\\ 
    Department of Biostatistics,\\ School of Global Public Health, New York University\\
    }
  \maketitle
} \fi

\if1\blind
{
  \bigskip
  \bigskip
  \bigskip
  \begin{center}
    {\LARGE\bf Regression Modeling of the Count Relational Data with Exchangeable Dependencies}
\end{center}
  \medskip
} \fi


\begin{abstract}
Relational data characterized by directed edges with count measurements are common in social science. Most existing methods either assume the count edges are derived from continuous random variables or model the edge dependency by parametric distributions. In this paper, we develop a latent multiplicative Poisson model for relational data with count edges. Our approach directly models the edge dependency of count data by the pairwise dependence of latent errors, which are assumed to be weakly exchangeable. This assumption not only covers a variety of common network effects, but also leads to a concise representation of the error covariance. In addition, the identification and inference of the mean structure, as well as the regression coefficients, depend on the errors only through their covariance. Such a formulation provides substantial flexibility for our model. Based on this, we propose a pseudo-likelihood based estimator for the regression coefficients, demonstrating its consistency and asymptotic normality. 
The newly suggested method is applied to a food-sharing network, revealing interesting network effects in gift exchange behaviors.
\end{abstract}

\noindent%
\vfill

\newpage
\spacingset{1.8}

\section{Introduction}
\label{sec1}

The modeling of relational data has garnered profound interests across various domains as it leads to a more comprehensive understanding of relationships in complex systems. 
Examples include deciphering brain connectivity maps \citep{Zhao2014differentialNetworks,
zhang2020mixed}, understanding friendship relationships, \citep{Moody2011Popularity}, and characterizing economic networks \citep{han2020individual}.
For such type of data, an important goal is to infer the mechanisms responsible for the observed relations, taking into account additional information on covariates within the relational structure.

\subsection{Background and motivation}

Consider relational data measured on pairs of $n$ nodes, where the directed edges may be assigned specific weights.
Efforts on modeling relational data have scattered in literature, with seminal examples including the social relations model \citep{Warner1979roundRobin,Wong1982roundRobin,kenny1984srm,
snijders1999srm}), which assumes normally distributed data and additive effects, and the row-column exchangeable model \citep{aldous1985exchangeability}.
In these modeling frameworks, the dependence structure within relational data are characterized by the latent variables.
To further incorporate with the possibly additional covariates information, \cite{Hoff2002Latent} developed latent space models, where they model the relational data as conditionally independent given the unobserved positions in social space of two nodes and the observed covariates that measure characteristics of the relational structure.
Aligned with the latent space models, the latent factor models \citep{Hoff2005mixedEffects,westveld2011mem} and additive and multiplicative effects (AME) models \citep{Hoff2013Likelihoods} have been proposed as parametric models based on the latent factors. 
Consequently, in contrast to the aforementioned approach, a more generalized model is imposed, referred to as dyadic regression models \citep{graham2020dyadic}, which operates without specifying a particular model form dependent on latent factors.
Beyond modeling a single network, there also exists a rich body of literature on temporal modeling of dynamic relational data \citep{suening2017network,kim2018review} and multiple relational data \citep{zhang2018network}.
Despite a handful of efforts on modeling relational data \citep{zhang2017estimating,Li2019predictNetwork,Hoff2021Additive,Le2022predictNetwork} as well as relational arrays \citep{Harris2011Longitudinal,Banerjee2013diffusiion,Marrs2023exchangeableErrors}, methods specifically designed for analyzing relational data with count observations remain limited.

Relational data characterized by weighted edges of count measurements are widely observed \citep{Krivitsky2012exponentialGraphs}, which we referred to as ``count relational data''.
Examples can be found in various domains such as coauthorship and citation networks \citep{ji2016coauthorship}, mobile phone communication networks \citep{Dong2012HMMs}, email exchange networks \citep{diesner2005exploration}, 
and transportation networks based on traffic counts \citep{Wang2009trafficCounts}.
A naive yet widely employed approach is to convert count edges to unweighted relational data with binary outcomes, which however may lead to information loss and spurious discoveries.
Among limited studies focused on the count relational data, \cite{Krivitsky2012exponentialGraphs} extended the exponential-family random graph models (ERGMs) to encompass the count outcomes.
As a pioneering method for estimating covariate effects on network data, ERGMs \citep{Holland1981exponentialGraphs,frank1986markov} relies on Markov chain Monte Carlo (MCMC) approximations to facilitate estimation that hinders its applicability for large relational data.
Also, the maximum likelihood estimator for ERGMs could be time-consuming \citep{caimo2011bayesian,schmid2017exponential} and has been found to be inconsistent under many network models \citep{Shalizi2013ConsistencyERGM}. 
Apart from ERGMs, there exists some efforts by utilizing multivariate counting processes to model counts of interactions when incorporating continuous time.
For instance, \cite{perry2013point} proposed a continuous-time model for dynamic data featuring directed counts of interactions, building upon event history analysis and assuming that no two interactions take place simultaneously. 
We refer to \cite{chen2023degree} for further discussions on such kind of dynamic models for continuous time relational data. 
Different from those works, we focus on a single network, employing a distinct approach to access the network dependencies.

\begin{figure}[ht!]
    \centering
    \includegraphics[scale=0.66]{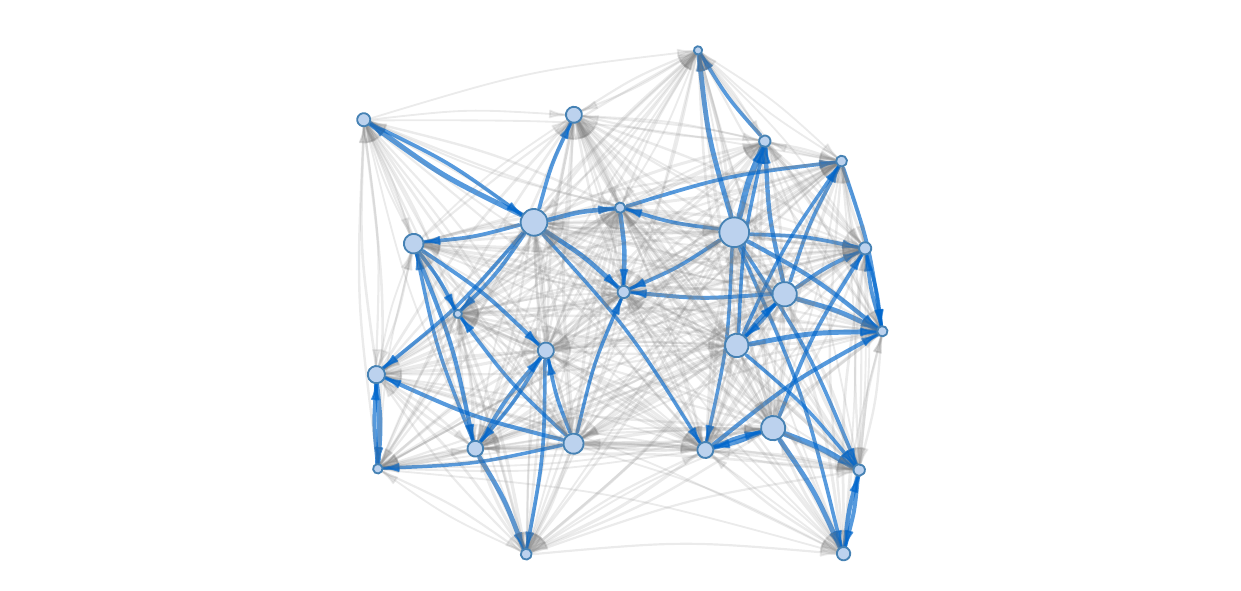} 
    \caption [Visualization of the food sharing network in Nicaragua.]{Plot of the food sharing network with the top-60 weighted count edges marked in blue.
    Larger node sizes indicate higher game harvest levels, and wider edges represent greater amount of gift exchanges between households.}
     \label{fig::food-sharing-visual-single} 
\end{figure}

Consider a motivating example of a food sharing network data collected by \cite{koster2014food}, which contains the number of transferred gifts among households over a yearlong period. Figure \ref{fig::food-sharing-visual-single} illustrates the food sharing network between households together with their game harvests. Gift transfers, particularly those of larger volume, mainly flow from households with substantial harvests to those with smaller yields. Households with lower game harvests tend to receive more gifts than those with higher harvests.
The authors adopt a Bayesian approach for analysis, noting that most existing methods are restricted to continuous response data, whereas the response variable in this case is a count.
To address the limitations of current methods discussed above, we aim to propose a modeling framework for \textbf{\textit{count relational data}} that accounts for \textbf{\textit{network dependencies}}, which enjoys theoretical guarantees and computational efficiency.

\subsection{Our contributions}

To address the challenges for modeling count relational data with the incorporation of node and edge covariates, we propose a latent multiplicative Poisson model. 
The main advantage is that our model facilitates the characterization of the edge dependency in relational data $\{y_{ij}\}_{1\leq i \neq j \leq n}$ directly through the latent errors $\{e_{ij}\}_{1\leq i \neq j \leq n}$, which yields an extra level of flexibility in the network structure that cannot be handled by ERGMs.
In addition, we avoid imposing any parametric specification on the latent errors.
This sets our work apart from the latent space model \citep{Hoff2002Latent} and other approaches \citep{Warner1979roundRobin,li2002unified} which enforce a parametric model of the latent factors.
The estimation of regression coefficients $\bbeta$ therefore needs to be tailored for the absence of full knowledge of likelihood. 
To this end, we employ the pseudo-maximum likelihood (PML, \cite{Gourieroux1984pseudoMLE}) approach.
We establish the asymptotic properties of estimated coefficients under network dependencies.
To lay the groundwork for the inference procedure, we further propose the estimation procedure of the covariance among $\{e_{ij}\}_{1\leq i \neq j \leq n}$.
Specifically, edge dependency within relational data can be empirically estimated by the frequencies of small subgraphs between two or three nodes \citep{opsahl2009clustering}.
When incorporating the proposed regression model, after getting the consistent estimation of coefficients employing the PML method, we demonstrate the consistency of the estimated covariance parameters via function of \emph{network moments} \citep{Zhang2022edgeworth}.
Our procedure is easily implemented and provides a computationally efficient approach for modeling count relational data compared to the existing methods that rely on MCMC approaches.

It is noteworthy that our proposed model differs from existing efforts that have utilized node-specific fixed effects to study the relational data \citep{graham2017econometric,dzemski2019empirical, chen2021nonlinear,zhang2023generalized}. 
We leverage a more general formulation that takes advantage of introducing the weak exchangeable \citep{Silverman1976weaklyExch} errors, which lends to us a concise representation of the covariance matrix.
Furthermore, our model sets itself apart from the model in \cite{graham2020dyadic} by relaxing the assumption that any pair of $(y_{ij},y_{kl})$ sharing at least one index in common are dependent, and accommodating edge covariates, which may not be encoded by the observable node information.

\subsection{Organization and notation}

Below, we introduce the latent multiplicative Poisson model on count relational data in Section \ref{sec2} and describe the structure of latent errors. 
In Section \ref{estbeta}, we present the PML estimation procedure and analyze the asymptotic normality of proposed estimator. 
With the consistently estimated covariance parameters of latent errors proposed in Section \ref{esteta}, we design valid inference procedures on the regression coefficients. 
Simulation studies are presented in Section \ref{sec4} to demonstrate the performance of our method. 
The application of our model to the food sharing network is given in Section \ref{sec5_food}, where we analyze social activities among households in Nicaragua. We conclude with discussions on future directions in Section \ref{sec6}. Technical proofs are given in the Supplementary Material.

\vspace{0.66em}
\noindent \textbf{Notation.}  The gradient and Hessian matrix of function $ g(\bx_{ij}^{\T}\bbeta)$ with respect to $\bbeta$ are represented by $\nabla g(\bx_{ij}^{\T}\bbeta)$ and $\nabla^2 g(\bx_{ij}^{\T}\bbeta)$, respectively. 
For ease of presentation, $\nabla g(\bbeta)$ and $\nabla^2 g(\bbeta)$ are used as their simplification. Denote the first and second derivatives of $g(\cdot) $ as $g'(\cdot) = \mathrm{d}g(z)/\mathrm{d}z $ and $g''(\cdot) = \mathrm{d}g'(z)/\mathrm{d}z$. 
The diagonal matrix with off-diagonals $\{a_{ij}\}$ is denoted as $\diag\{a_{ij}\}$. 
Unless specified otherwise, we write the $\ell_2$-norm of a vector as $\|\ba\|$. 
Let $\xrightarrow{d}$, $\xrightarrow{p}$ and $\stackrel{\mathbb{P}_{\bbeta}}{\rightarrow}$ represent the convergence in distribution, in probability, and in probability given $\bbeta$, respectively. 
Throughout the rest of the paper, $\{\cdot_{ij}\}$ is used to denote the set of variables indexed by $i,j$, \ie $\{\cdot_{ij}\}_{1\leq i \neq j \leq n}$, for clarity.

\section{Problem set up}
\label{sec2}

\subsection{Latent multiplicative Poisson model}
\label{secModel}

Consider observing the count outcomes $\{y_{ij}\}$ among $n$ nodes indexed by $1,\ldots,n$ as well as the covariates $\{\bx_{ij} : \bx_{ij} \in \mathbb{R}^p\}$, where self-loops are excluded as node's interaction with itself is not of our interest. 
As discussed in Section \ref{sec1}, existing models that focus on binary outcomes \citep{Hoff2002Latent} or with additive errors \citep{minhas2019inferential,Le2022predictNetwork,Marrs2023exchangeableErrors} do not naturally handle count data.  
Inspired by the conditional autoregressive models on counting process 
\citep{Zeger1988Count,Davis2000Count,Diggle2002Count}, we introduce the following latent multiplicative Poisson model:
\begin{align}
    y_{ij}|\lambda_{ij}\sim \text{Poisson}\left(\lambda_{ij} := g(\bx_{ij}^{\T}\bbeta)e_{ij}\right),
    \label{model1}
\end{align}
where $\bbeta = (\beta_1, \beta_2, \ldots, \beta_p)^{\T} \in \mathbb{R}^p$ is the vector of regression coefficients, $g: \mathbb{R} \rightarrow (0, \infty)$ serves as the link function, and the latent error $e_{ij}$ admits unit mean such that  $\mathbb{E}(\lambda_{ij}) = g(\bx_{ij}^{\T}\bbeta)$. 
The choice of $g(\cdot)$ is pre-specified, such as logistic function, arc-cotangent function, and exponential function, to name a few. 
In model \eqref{model1}, the dependence across observations are modeled via latent errors $\{e_{ij}\}$, whose distribution does not need to be specified and therefore leads to great flexibility of our model.  

The multiplicative nature of $\lambda_{ij}$ with respect to the regression component and the latent error benefits in two ways. 
First, it facilitates an analogous way to define the residual of count relational data without specifying a stringent parametric model. 
Second, it establishes a direct connection between the data dependence represented by the covariance of $\{y_{ij}\}$, and the dependence among latent errors characterized by the covariance of $\{e_{ij}\}$, as demonstrated in \eqref{CovYij_1}.
This cannot be achieved via an additive model for $\{\lambda_{ij}\}$ as the positivity of $\{\lambda_{ij}\}$ is not easily aligned with traditional assumption on $\mathbb{E}(e_{ij})=0$. 
Under model \eqref{model1}, we have 
\begin{align}
    \label{CovYij_1}
    \text{Cov}(y_{ij},y_{i'j'})=g(\bx_{ij}^{\T}\bbeta)g(\bx_{i'j'}^{\T}\bbeta)\text{Cov}(e_{ij},e_{i'j'}), \text{ for } i \neq i' \text{ or } j \neq j';
\end{align}    
and
\begin{align}
    \text{Var}(y_{ij}) = g^2(\bx_{ij}^{\T}\bbeta) \text{Var}(e_{ij}) + g(\bx_{ij}^{\T}\bbeta).
    \label{VarYij}
\end{align}
Let $\xi_{ij}= y_{ij} \{g(\bx_{ij}^{\T}\bbeta)\}^{-1}$, we have $\mathrm{Cov}(e_{ij},e_{i'j'}) = \mathrm{Cov}(\xi_{ij},\xi_{i'j'})$ by \eqref{CovYij_1}, which hints a natural covariance estimator of $\{e_{ij}\}$ in Section \ref{esteta}.
Moreover, covariance terms of $\{e_{ij}\}$ in \eqref{CovYij_1}, namely $\mathrm{Cov}(e_{ij},e_{ji}), \mathrm{Cov}(e_{ij},e_{ik}), \mathrm{Cov}(e_{ij},e_{kj})$, and $\mathrm{Cov}(e_{ij},e_{ki})$ represent the commonly-encountered network effects \citep{cranmer2014reciprocityEffect,du2024optimal}, which we refer to as the reciprocity effect, same sender effect, same receiver effect, and sender-receiver effect, respectively. 
Such dependency structure is denoted as the social relations covariance model in \cite{Hoff2021Additive} under parametric model assumptions.

\subsection{Weakly exchangeable latent errors}
\label{EE}

We are now in position of modeling the latent errors $\{e_{ij}\}$, which lends a concise representation of the dependence among edges $\{y_{ij}\}$. 
Instead of imposing parametric assumptions, we only assume that the latent errors are weakly exchangeable \citep{Silverman1976weaklyExch}. 
An array $\bz=\{z_{i,j}\}$ is called weakly exchangeable if $\{z_{i,j}\}\stackrel{d}{=}\{z_{\pi(i), \pi(j)}\}$ for any simultaneous permutation $\pi(\cdot)$ of both the row and column labels.
Such an assumption is desirable for relational data as it is of great interest to relate the outcomes involving node $i$ as a sender to that involving $i$ as a receiver.
In this paper, we focus on the dissociated weakly exchangeable array to model latent variables, where any random variables within the array are independent whenever their indexing sets are disjoint.


A major appeal of the weak exchangeability is the concise parametrization of the covariance matrix of $\bfe =(e_{12}, e_{13}, \ldots, e_{n-1n})^{\T}\in \mathbb{R}^{n^2-n}$, denoted by $\bOmega_e := \mathbb{E}\big\{\big(\bfe - \mathbb{E}(\bfe)\big)\big(\bfe - \mathbb{E}(\bfe)\big)^{\T}\big\}$. 
In fact, six unique parameters are sufficient to parameterize the $O(n^4)$ entries of $\bOmega_e$, since there exist six distinguishable configurations of pairs drawn from $\{e_{ij}\}$ with unlabeled nodes \citep{Hoff2021Additive,Marrs2023exchangeableErrors}. 
Specifically, the diagonal of $\bOmega_e$ admits $\eta_1=\mathrm{Var}(e_{ij})$, while off-diagonals $\mathrm{Cov}(e_{ij}, e_{ji}), \mathrm{Cov}(e_{ij}, e_{il}), \mathrm{Cov}(e_{ij}, e_{kj})$ and $\mathrm{Cov}(e_{ij}, e_{ki})$ are represented by $\eta_2, \eta_3, \eta_4$ and $\eta_5$, respectively.
Additionally, we assume $\mathrm{Cov}(e_{ij}, e_{kl})=0$ for $\{i,j\}\cap\{k,\ell\}=\emptyset$ according to the dissociated array assumption \citep{Silverman1976weaklyExch}. 
With such a parameterization, the multiplicities of $\eta_1$ to $\eta_5$ in each row/column of $\bOmega_e$ are $1, 1, n-2, n-2, 2n-4$, respectively, while remaining entries are zero. Denote  $\bfeta:=(\eta_1, \eta_2, \eta_3, \eta_4, \eta_5)^{\T}$. 
For the non-negative definiteness of $\bOmega_e$, $\bfeta$ should fall in the following parameter space:
\begin{align*} 
    \mathcal{M}^n(\eta_1,\eta_2,\eta_3,\eta_4,\eta_5) = &
    \{\mathbb{R}^5:  \eta_5 \geq - (\eta_3 + \eta_4)/2 - (\eta_2 + \eta_1)/ (2n - 4),~ -\eta_1 \leq \eta_2 \leq \eta_1, \\
    & \eta_5 \leq (\eta_1 + \eta_2 - \eta_3 - \eta_4) / 2,~
    \eta_5 \geq (- \eta_1 + \eta_2 + \eta_3 + \eta_4) / 2,~  \\
     & \eta_1 \geq 0,~ \text{and}~ \{(n-3)(\eta_3+\eta_4)-2\eta_5+2\eta_1\}^2 \geq \iota + \kappa \},
\end{align*}
where $\iota = (\eta_4^2+\eta_3^2)(n^2-2n+1) + 4\eta_5^2(n^2-6n+9) + 2\eta_3\eta_4(1-n^2+2n)$ and $\kappa = \eta_2\eta_5(8n- 24) + (\eta_3+\eta_4)\eta_5(12-4n) + 4\eta_2\{\eta_2-(\eta_3+\eta_4)\}$. 
Details are relegated to the Supplementary Material. 
In practice, this helps to establish an valid estimation of $\bfeta$ to draw inference on $\bbeta$, which will be discussed in Section \ref{esteta}.
    
It is common to assume dependence between edges with sharing nodes, such as in the social relations model \citep{Warner1979roundRobin,Wong1982roundRobin,kenny1984srm,Gill2001srm}, the conditionally independent dyad model \citep{chandrasekhar2016cid,graham2020cid}, and the random-effects model \citep{gelman2006ref,westveld2011mem,aronow2015cluster,graham2021minimax}.
For relational data, those dependencies characterized by covariance terms as defined in Section \ref{secModel}, representing the network effects between edges sharing common nodes.
In our work, we assume at least one of $\{\eta_3,\eta_4,\eta_5\}$ is nonzero.
In practice, the weakly exchangeable array can be easily generated from a variety of widely-used models, as exemplified in Example \ref{ep1}.
The proof of its weak exchangeability is given in the Supplementary Material.

\begin{example}[Weakly exchangeable errors from linear mixed effects models]
    \label{ep1} 
    \emph{Consider $e_{ij} = C (a_i + b_j + \gamma _{(ij)} + \epsilon_{ij})$. 
    Here, $(a_i,b_i)^{\T}$ is bivariate truncated normal with location $\bmu_0=(\mu_{a_0},\mu_{b_0})^{\T}$, covariance $\Sigma_{ab} = [(\sigma^{2}_{a_0}, \rho_{0}\sigma_{a_0}\sigma_{b_0})^{\T}; (\rho_{0}\sigma_{a_0}\sigma_{b_0}, \sigma^{2}_{b_0})^{\T}]$ and truncation parameters $\bv_0,\bu_0$;     
    $\gamma _{(ij)} = \gamma _{(ji)} \sim \mathrm{truncN}(\mu_{\gamma_0},\sigma^{2}_{\gamma_0},v_{\gamma_0},u_{\gamma_0})$ and $\epsilon_{ij} \sim \mathrm{truncN}(\mu_{\epsilon_0}, \sigma^{2}_{\epsilon_0},v_{\epsilon_0}, u_{\epsilon_0})$. 
    The normalization constant $C = (\mu_{a} + \mu_{b} + \mu_{\gamma} + \mu_{\epsilon})^{-1}$, where $\mu_{a}$, $\mu_{b}$, $\mu_{\gamma}$, and $\mu_{\epsilon}$ are the means after truncation corresponding to independent $a_i, b_j, \gamma_{(ij)}$ and $\epsilon_{ij}$.
    }  \end{example}

\section{Estimation and inference on regression coefficients} 
\label{estbeta}  

As the primary task to analyze the relational data under model \eqref{model1}, we estimate the regression coefficients $\bbeta = (\beta_1, \beta_2, \ldots, \beta_p)^{\T}$ by utilizing the pseudo-likelihood, from which the estimator's asymptotic distribution is carefully established to draw inference on $\bbeta$. 

The dependence among $\{e_{ij}\}$ imposes difficulty to estimate the regression coefficients, since integrating out the latent variables requires the knowledge of the joint distribution of $\{e_{ij}\}$. However, in our model, the weak exchangeability of $\{e_{ij}\}$ is a mild assumption, where the joint distribution of errors does not need to be specified. 
To address this, we employ the pseudo-maximum likelihood (PML, \cite{Gourieroux1984pseudoMLE}) method.
For data comes from a linear exponential family, PML necessitates only the accurate specification of the mean structure of data generating processes to ensure consistent estimation of coefficients. 
Though it can not easily work with dependence as pointed by \cite{Besag1975PML}, the observations are independent conditional on the error terms. We therefore work on the conditional pseudo-likelihood since our target $\bbeta$ is on the mean structure of model \eqref{model1}. Enlightened by this approach, we maximize a likelihood function associated with a family of probability distributions. 
Conditional on latent errors $e_{ij}$, the distribution of $\{y_{ij}\}$ is $\prod^{n}_{i = 1} \prod^{n}_{\substack{j = 1; j \neq i}} \exp(-\lambda_{ij})(\lambda_{ij})^{y_{ij}}(y_{ij}!)^{-1}$ and 
therefore suggests an objective function under model \eqref{model1} as
\begin{align}
    \ell_n (\bbeta) = (n^2-n)^{-1} \sum \limits_{\substack{i,j=1\\ i \neq j}}^{n} \Big[y_{ij} \text{log}\{g(\bx_{ij}^{\T} \bbeta)\} - g(\bx_{ij}^{\T} \bbeta)\Big].
    \label{PoisObjFun}
\end{align}
Maximizer to $\ell_n (\bbeta)$, $\wh \bbeta_n$ is an estimator of $\bbeta$. To facilitate our derivation, we introduce a few notations. For relational data with $n$ nodes, let $S_{1,n} := \{\{(i,j),(i,j)\}: \text{distinct~} i, j \in [n]\}$; $S_{2,n} := \{\{(i,j),(j,i)\}: \text{distinct~} i, j \in [n]\}$; $S_{3,n} := \{\{(i,j), (i,k)\}: \text{distinct~} i, j, k \in [n]\}$; $S_{4,n} := \{\{(i,j), (k,j)\}: \text{distinct~} i, j, k \in [n]\}$; and $S_{5,n} := \{\{(i,j), (k,i)\}: \text{distinct~} i, j, k \in [n]\} \cup \{\{(i,j), (j,k)\}: \text{distinct~} i, j, k \in [n]\}$. Now we impose the following conditions to draw inference on the coefficients.
\begin{assumption}
    \label{as1}
    \begin{enumerate}[(a)]
        For the relational data $\{y_{ij}\}$ generated from model \eqref{model1}, we assume
        \item \label{consistency_1} the limit of $~|S_{m,n}|^{-1} \sum_{(ij, kl) \in S_{m,n}} [\nabla g(\bx^{\T}_{ij} \bbeta)]^{\T} \nabla g(\bx^{\T}_{kl} \bbeta)$ exists for $m \in \{1, 2, 3, 4, 5\}$; 
        \item \label{covariance_1} the limit of $(n^2-n)^{-1} \sum\nolimits_{\substack{i,j=1, i \neq j}}^{n} [\nabla g(\bx^{\T}_{ij} \bbeta)]^{\T} \nabla g(\bx^{\T}_{ij} \bbeta) \{g(\bx^{\T}_{ij} \bbeta)\}^{-1}$ exists and it is invertible, denoted by $\mathbf{J}$;
        \item \label{covariance_2} the limit of $(n^2-n)^{-1} \| \sum\nolimits_{\substack{i,j \in [n]; i \neq j}} \bx_{ij} \bx^{\T}_{ij}\|$ and $(n^2-n)^{-1} \| \sum\nolimits_{\substack{i,j= \in [n]; i \neq j}} \bx_{ij} \bx^{\T}_{ij} \bx_{ij}\|$ exist, both $[g''(\cdot)\{g(\cdot)\}^{-1}]'$ and $\{\text{log}(g(\cdot))\}'' \{g(\cdot)\}^{-1}$ are $L_2$ integrable;
        \item \label{CLT_1} moment condition on $\{e_{ij}\}$: $\|e_{ij}\|_4 = \mathbb{E}(|e_{ij}|^4)^{1/4} < L < \infty$, for $i,j \in [n]$;
        \item \label{CLT_2} for any $\mathbf{t} \in \mathbb{R}^p$, there exists $L_0 < \infty$ such that \text{sup}$_{i,j} \mathbf{t}^{\T} \nabla g(\bx^{\T}_{ij} \bbeta) = L_0 < \infty$.
    \end{enumerate}
\end{assumption}

Assumption \ref{as1} establishes the regularity conditions for deriving the asymptotic behavior of $\wh \bbeta_n$. 
First, Conditions \eqref{consistency_1} and \eqref{covariance_1} impose mild assumptions on the gradient of the link function, extensively employed in Poisson model literature \citep{Davis2000Count,davis2016theory}.
Condition \eqref{covariance_2} incorporates summability assumptions on the covariates and the integrability constraints on $g(\cdot)$, both of which are widely used in regression contexts \citep{phillips1999linear}.
It could be easily satisfied by linear functions and other functions discussed in Section \ref{secModel}. 
Together, these three conditions ensure the consistency of $\wh \bbeta_n$ and regulate the limiting behavior of the asymptotic covariance of $\wh \bbeta_n$. 
Second, the asymptotic normality of $\wh \bbeta_n$ is guaranteed by conditions \eqref{CLT_1} and \eqref{CLT_2}. 
Condition \eqref{CLT_1} aligns with assumptions made by \cite{Lumley2003sparseCorrelations} and \cite{Marrs2023exchangeableErrors}.
The moment condition in \eqref{CLT_1} exhibits a higher degree of flexibility when compared to those found in high-dimensional generalized linear models \citep{tian2022transfer}, which typically impose a light tail assumption on random noises. 
Indeed, there exists a broad spectrum of distributions for $\{e_{ij}\}$ satisfying Condition \eqref{CLT_1}, such as sub-Gaussian and sub-exponential families. 
Finally, Condition \eqref{CLT_2} can be met by employing suitable encoding techniques to ensure that $\bX$ remains within a compact domain and is trivially satisfied in the case of fixed design.

Now we are ready to formally provide the inference procedure of $\bbeta$. The asymptotic property of $\wh \bbeta_n$ is summarized in Theorem \ref{co1}. 
Here we focus on the population version with respect to the true asymptotic covariance matrix.

\begin{thm}
    \label{co1}
    Assume the data is generated from model \eqref{model1}. Under Assumption \ref{as1}, the pseudo-likelihood estimate $\wh \bbeta_n$ is asymptotically normal:
    \begin{align*}
         \sqrt{n}(\wh \bbeta_n - \bbeta) \xrightarrow{d} \mathcal{N} (\textbf{0}, \mathbf{J}^{-1}\mathbf{L}\mathbf{J}^{-1}),
    \end{align*}
    with $\mathbf{J} = \lim_{n \to \infty} \mathbf{J}_n$, $\mathbf{L} = \lim_{n \to \infty} \mathbf{L}_n$, where $\mathbf{J}_n = (n^2 - n)^{-1}  \nabla g(\bbeta) \bSigma^{-1}_0 \nabla g(\bbeta)^{\T} $, $\mathbf{L}_n = (n^3-n^2)^{-1} \nabla g(\bbeta) \bSigma^{-1}_0 \bOmega_0 \bSigma^{-1}_0 \nabla g(\bbeta)^{\T} $. Within $\mathbf{J}_n$ and $\mathbf{L}_n$, $\bSigma_0$ denotes the variance matrix associated with chosen linear exponential family, and let $\bOmega_0 = \bOmega(\bX, \bbeta)$ be the covariance of $\bY$. Given design matrix $\bX$, we have $\bSigma_0  = \bSigma(\bX, \bbeta) = \diag\{g{(\bx_{ij}^{\T}\bbeta)}\} \in \mathbb{R}^{(n^2-n) \times (n^2-n)}$.
\end{thm}

Theorem \ref{co1} paves a road for drawing inference on $\bbeta$.
As discussed in Section \ref{EE}, we focus on the setting with network dependence where at least one of $\{\eta_3,\eta_4,\eta_5\}$ is nonzero.
Due to the complex network dependencies, the likelihood of $y_{ij}$ cannot be expressed as the sum of independent random variables.  
As a result, the standard central limit theorems and Lindeberg's condition are not directly applicable. 
To address this issue, one might consider leveraging U-statistics techniques to induce a summation of independent random variables \cite{graham2020dyadic,graham2020cid}.
However, such an approach requires additional assumptions on the distribution of $\{y_{ij}\}$. 
For example, \cite{graham2020dyadic} assumes exchangeability of $\{y_{ij}\}$, and does not naturally accommodate edge covariates.
In contrast, we assume exchangeability only on the error terms in the regression model.
Moreover, \cite{graham2020dyadic} assumes dependence between any pair $(y_{ij}, y_{kl})$ sharing at least one index, whereas our model does not impose this requirement. 
In fact, the unknown dependency structure can lead to indeterminate degenerate status of U-statistics under network setting \citep{du2024optimal}.  
To overcome these challenges, we employ a key lemma from \cite{Bolthausen1982mixingCLT}, which provides a sufficient condition for establishing the asymptotic normality of a sequence of measures based on the standard normal characteristic function. The proof of Theorem \ref{co1} is given in the Supplementary Material.

The existence of $\mathbf{J}_n$ and $\mathbf{L}_n$ are guaranteed by conditions \eqref{consistency_1} and \eqref{covariance_1}. 
By replacing $\bbeta$ and $\bfeta$ by $\wh \bbeta_n$ and $\wh \bfeta$ in $\mathbf{J}_n$ and $\mathbf{L}_n$ to get $\mathbf{\wh J}_n$ and $\mathbf{\wh L}_n$, the asymptotic covariance matrix can be estimated through $\mathbf{\wh J}_n^{-1} \mathbf{\wh L}_n \mathbf{\wh J}_n^{-1}$.
We will demonstrate its consistency in Section \ref{esteta}.
The following Example \ref{ep_exp_link} provides an illustration of Theorem \ref{co1} with exponential link function, which yields a simple form of the asymptotic covariance matrix.
  
\begin{example} 
    \label{ep_exp_link}
    \emph{Under the setting in Theorem \ref{co1}, choose exponential link function and denote $\bLambda$ as $\diag\{g{(\bx_{ij}^{\T}\bbeta)}\}$. The asymptotic variance of the estimator $\wh \bbeta_n$ is given below:
    \begingroup
    \allowdisplaybreaks
    \begin{align*}
        \mathbf{J}^{-1}\mathbf{L}\mathbf{J}^{-1}
        =&  \lim_{n \to \infty} n \Big[\sum^n_{i\neq j}\{\bx_{ij}   \bx_{ij}^{T}  g{(\bx_{ij}^{\T}\bbeta) }\}\Big]^{-1}  \Big(\bX   \bOmega_0     \bX^{T}\Big) \Big[ \sum^n_{i\neq j}\{\bx_{ij}   \bx_{ij}^{T}  g{(\bx_{ij}^{\T}\bbeta) }\}\Big]^{-1},
    \end{align*}
    \endgroup
    where $\bOmega_0 = \text{Cov}(y_{ij})$ is the covariance matrix of $\bY$, which is fully specified using $\bfeta$ from the weakly exchangeable latent errors by \eqref{CovYij_1} and \eqref{VarYij}. }
\end{example}

\section{Estimation on covariance parameters {\protect\boldmath $\eta$}} 
\label{esteta}  

In this section, we construct a consistency estimator of the asymptotic covariance matrix in Theorem \ref{co1}. This involves the estimation of $\bfeta$, which serves as the cornerstone for inferring $\bbeta$. Recall that $\eta_1$ is the variance of error terms while $\eta_2$ through $\eta_5$ are the covariance terms in $\bOmega_e$, and $S_{1,n}$ through $S_{5,n}$ represent the sets of pairs of links corresponding to $\eta_1$ through $\eta_5$, as defined in Section \ref{estbeta}. First, we develop the estimation procedure of the covariance terms. By \eqref{CovYij_1}, given the knowledge of $\{\xi_{ij}\}$, we consider the moment estimator of $\eta_2$ through $\eta_5$ as follows:
\begingroup
\allowdisplaybreaks
\begin{align*} 
    \overline \eta_2 =&~ |S_{2,n}|^{-1} \sum^n_{i\neq j}   \xi_{ij}   \xi_{ji} - |S_{2,n}|^{-2} \Big( \sum^n_{i\neq j}   \xi_{ij} \Big) \Big( \sum^n_{i\neq j}   \xi_{ji} \Big) , \text{ for } \{(i,j),(j,i)\} \in S_{2,n}, \\ 
    \overline \eta_3 =&~ |S_{3,n}|^{-1} \sum^n_{i\neq j \neq l}   \xi_{ij}   \xi_{il} - |S_{3,n}|^{-2}  \Big( \sum^n_{i\neq j }   \xi_{ij} \Big) \Big( \sum^n_{i\neq l }   \xi_{il} \Big) , \text{ for } \{(i,j),(i,l)\} \in S_{3,n}, \\ 
    \overline \eta_4 =&~ |S_{4,n}|^{-1} \sum^n_{i\neq j \neq k}   \xi_{ij}   \xi_{kj} - |S_{4,n}|^{-2}  \Big( \sum^n_{i\neq j }   \xi_{ij} \Big) \Big( \sum^n_{k\neq j} \xi_{kj} \Big), \text{ for } \{(i,j),(k,j)\} \in S_{4,n}, \\ 
    \overline \eta_5 =&~ |S_{5,n}|^{-1} \sum^n_{i\neq j \neq k}   \xi_{ij} (  \xi_{ki} +   \xi_{jk}) - 2\cdot |S_{5,n}|^{-2} \Big( \sum^n_{i\neq j }   \xi_{ij} \Big) \Big(\sum^n_{k\neq i}   \xi_{ki} + \sum^n_{k\neq j}   \xi_{jk}\Big),
\end{align*}
\endgroup
for $\{(i,j),(k,i)\} \in S_{5,n}$ and $\{(i,j),(j,k)\} \in S_{5,n}$. Replacing $\{\xi_{ij}\}$ in $\overline \eta_2, \overline \eta_3, \overline \eta_4$, and $\overline \eta_5$ by $\{\wh \xi_{ij}\}$ gives us $\wh \eta_2, \wh \eta_3, \wh \eta_4$, and $\wh \eta_5$. 

For the variance term $\eta_1 = \text{Var}(\xi_{ij}) - \{g(\bx_{ij}^{\T}\bbeta)\}^{-1}$, a natural estimator is $\overline \eta_{1,*} = |S_{1,n}|^{-1} \sum^n_{i\neq j} \xi^2_{ij} - |S_{1,n}|^{-2} (\sum^n_{i\neq j} \xi_{ij})^2 - |S_{1,n}|^{-1} \sum^n_{i\neq j} \{g(\bx_{ij}^{\T} \bbeta)\}^{-1}$, given the knowledge of $\bbeta$. In practice, replacing $\bbeta$ by $\wh \bbeta_n$ obtained from \eqref{PoisObjFun} gives:
\begin{equation} 
    \begin{aligned}
    \wh \eta_{1,*} =& |S_{1,n}|^{-1} \sum^n_{i\neq j} \wh \xi^2_{ij} - |S_{1,n}|^{-2} \Big(\sum^n_{i\neq j} \wh \xi_{ij}\Big)^2 - |S_{1,n}|^{-1} \sum^n_{i\neq j} \{g(\bx_{ij}^{\T} \wh \bbeta_n)\}^{-1}.
    \label{eta1}
   \end{aligned}
\end{equation} 
However, the estimator in \eqref{eta1} does not necessarily guarantee the positivity of the variance term, which may not lead to a legitimate $\wh \bOmega_e$. To circumvent that difficulty, we refine the estimator using a hybrid procedure, which proceeds by first applying \eqref{eta1}, and then modify $\wh \eta_{1,*}$ using a $k$-shorth estimator \citep{Pensia2019entangled}. The $k$-shorth estimator outputs the center of the shortest interval containing at least $k$ points. While the traditional shorth estimator uses $k = N/2$ for sample size $N$, the estimator in \cite{Pensia2019entangled} considered $k = c\log(N)$ for tunning parameter $c$, which provides the guaranteed optimality.  

By \eqref{VarYij}, we have $\eta_1 = \mathbb{E}(\xi^2_{ij}) - 1 - \{g(\bx_{ij}^{\T}\bbeta)\}^{-1}$ for distinct $i, j \in [n]$. 
Let $\wh \xi_{ij}= {y_{ij}} \{g(\bx_{ij}^{\T}\wh \bbeta_n)\}^{-1}$ and $\wh \zeta_{ij} = \wh \xi^2_{ij} -  1 - \{g(\bx_{ij}^{\T}\wh \bbeta_n)\}^{-1}$, then the estimator in \eqref{eta1} can be rewritten as $(n^2 - n)^{-1} \sum^n_{i\neq j} \wh \zeta_{ij}$. We consider the $n^2-n$ elements in $\{\wh \zeta_{ij}\}$ to be the total points we apply the shorth estimation procedure on. Then we control the size of the $k$-shorth interval by letting $k = c\log(N)$ for $N=n^2 - n$, and take only the valid intervals with a positive center to constrain the estimator to give us a positive estimate of $\eta_1$, denoted as $\wh \eta_{1, +}$. The proposed hybrid estimation procedure outputs the estimate in \eqref{eta1} when it is greater than zero; otherwise, it outputs the positive $k$-shorth estimate. The hybrid estimator could be represented by $\wh \eta_1 = \wh \eta_{1,\mathrm{hybrid}} = \wh \eta_{1, +} \cdot \mathbb{I}(\wh \eta_{1,*} \leq 0) + \wh \eta_{1,*}\cdot \mathbb{I}(\wh \eta_{1,*} > 0).$

Let $\wh \bfeta = (\wh \eta_1, \wh \eta_2, \wh \eta_3, \wh \eta_4, \wh \eta_5)^{\T}$ denote the estimator of $\bfeta$. Theorem \ref{thm2} below establishes the consistency of $\wh \bfeta$. When the number of nodes goes to infinity, $\wh \bfeta$ will fall in the parameter space discussed in Section \ref{EE} by its consistency. 
\begin{thm}
    \label{thm2}
    Under the assumptions of Theorem \ref{co1}, the covariance of weakly exchangeable errors are consistently estimated in the sense that $\wh \eta_i - \eta_i \xrightarrow{p} 0$ for $i \in \{1,2,3,4,5\}$.
\end{thm}

By Theorem \ref{thm2}, we consider the covariance estimator of $\bOmega_0$ in Theorem \ref{co1} takes the form $\wh \bOmega_0 = \widehat{\text{Cov}}(y_{ij})$, where
$\widehat{\text{Var}}(y_{ij}) = g(\bx_{ij}^{\T}\wh \bbeta_n)^2 \wh \eta_1 + g(\bx_{ij}^{\T}\wh \bbeta_n)$;
$\widehat{\text{Cov}}(y_{ij}, y_{ji}) = g(\bx_{ij}^{\T}\wh \bbeta_n)g(\bx_{ji}^{\T}\wh \bbeta_n)\wh \eta_2$;
$\widehat{\text{Cov}}(y_{ij}, y_{il}) = g(\bx_{ij}^{\T}\wh \bbeta_n)g(\bx_{il}^{\T}\wh \bbeta_n)\wh \eta_3$;
$\widehat{\text{Cov}}(y_{ij}, y_{kj}) = g(\bx_{ij}^{\T}\wh \bbeta_n)g(\bx_{kj}^{\T}\wh \bbeta_n)\wh \eta_4$;
$\widehat{\text{Cov}}(y_{ij}, y_{jk}) =  g(\bx_{ij}^{\T}\wh \bbeta_n)g(\bx_{jk}^{\T}\wh \bbeta_n)\wh \eta_5$,
and ${\text{Cov}}(y_{ij}, y_{kl}) = 0$ by the dissociated assumption introduced in Section \ref{EE}. Note that to obtain the asymptotic properties of $\wh \bbeta_n$, $\bOmega_0$ needs to be invertible. The following proposition shows that the consistent estimators of $\eta_i$'s provide us a positive define $\wh \bOmega_0$ as the estimator of covariance matrix of $\bY$.

\begin{proposition} \emph{Given $\wh \bOmega_e$ is positive semi-define, $\wh \bOmega_0$ is a positive define matrix.}\label{prop31} \end{proposition}

Finally, we show the asymptotic covariance matrix in Theorem \ref{co1} can be consistently estimated in practice.

\begin{thm}
    \label{var_consis}
    Under the assumptions of Theorem \ref{co1}, $\mathbf{\wh J}_n^{-1} \mathbf{\wh L}_n \mathbf{\wh J}_n^{-1}$ is consistent for $\mathbf{J}^{-1}\mathbf{L}\mathbf{J}^{-1}$.
\end{thm}

Combining Theorem \ref{var_consis} and Theorem \ref{co1} leads to a formal inference procedure of $\bbeta$. For example, for each $\ell \in [p]$, denote $\wh \sigma^2_\ell$ the $\ell$th diagonal entry of $n^{-1}\mathbf{\wh J}_n^{-1} \mathbf{\wh L}_n \mathbf{\wh J}_n^{-1}$, a $100(1-\alpha)\%$ confidence interval for the $\ell$th entry of $\bbeta$, denoted by $\bbeta_\ell$, is given by $\big[\wh \bbeta_\ell - \wh \sigma_\ell \bPhi^{-1}(1-\alpha/2), \wh \bbeta_\ell + \wh \sigma_\ell \bPhi^{-1}(1-\alpha/2)\big]$, where $\bPhi(\cdot)$ is the cumulative distribution function of standard normal distribution.

\section{Simulation studies}
\label{sec4}

In this section, we evaluate the numerical performance of the proposed method for count relational data and compare it with other benchmarks. 
We illustrate the validity of our inference framework under the weakly exchangeable error setting in Section \ref{EE}. 
Since few models have been studied for count relational data and even fewer for the dependence structure introduced in this paper, we examine the performance of 95\% confidence interval coverage probability among the three methods: 
(1) {\tt Our model}: the inference procedure proposed in our work;
(2) {\tt Naive}: the inference procedure assuming no edge dependencies;
(3) {\tt Oracle}: the inference procedure given the true covariance matrix of error terms. 
The oracle result with known value of $\bfeta$ serves as a benchmark, while the naive method assumes observations $\bY$ are marginally independent.
The results of the naive method will demonstrate the necessity of involving the dependency structure of relational data in the inference procedure.
The following model is employed to generate count relational data:
\begin{align}
    y_{ij} \sim \text{Poisson}\Big(\exp\big\{\beta_1 x_{1ij} + \beta_2 x_{2i}x_{2j} + \beta_3 |x_{3i} - x_{3j}| + \beta_4 x_{4ij} \big\} e_{ij}\Big).
    \label{simuModel}
\end{align}
We vary the number of nodes $n \in \{20 ,50, 100, 150\}$, fix $\bbeta = (1, -0.5, -0.5, -1)^{\T}$, and independently draw $x_{1ij}\sim N(2,1)$, $x_{2i}\sim \mathrm{Bernoulli}(0.6)$, $x_{3i}\sim N(1,1)$, and $x_{4ij}\sim N(1,1)$.
We apply settings in Example \ref{example::_gen_err_truncNormal} to generate $\{e_{ij}\}$.
Under each realization of $\bX$, we simulate 1,000 error terms to calculate the empirical coverage probability, and repeat the experiment 15 times to evaluate the variation in the 95\% confidence interval coverage of the three competing methods.

\begin{example}[Generating mechanism for error terms]
\label{example::_gen_err_truncNormal}
~
\begin{enumerate} [(i)] 
    \item
    \label{gen1}
    Independent and identically distributed errors, where $e_{ij} \sim \mathrm{truncN}(-7,1,0,\infty)$, and are normalized to satisfy the model assumption of positive error terms with unit mean.  
    \item 
    \label{gen2}
    Dependent errors with weakly exchangeable structure under Example \ref{ep1}, where $(a_i,b_i)^{\T}$ is generated with $\bmu_0=(-1, 1)^{\T}$, $\Sigma_{ab} = (1, 0.5;0.5,1)$, $\bv_0=(0, 0)^{\T}$, and $\bu_0=(\infty, \infty)^{\T}$; $\gamma _{(ij)} = \gamma _{(ji)} \sim  \text{truncN}(0,1,0,\infty)$, and $\epsilon_{ij} \sim \text{truncN}(1, 6,0,\infty)$. 
\end{enumerate}
\end{example}

By construction, we have $\bfeta = (1.1, 0, 0, 0, 0)$ under setting \eqref{gen1}, and $\bfeta = (13, 2, 7, 2, 0.4) \times 10^{-2}$ under setting \eqref{gen2}.
Although setting \eqref{gen1} falls outside the scope of our primary interest due to the absence of edge dependencies, we can still estimate the asymptotic covariance matrix of $\wh \bbeta$ by $n^{-1}\mathbf{\wh J}_n^{-1} \mathbf{\wh L}_n \mathbf{\wh J}_n^{-1}$. 
This serves as a configuration which admits the assumption of independence among edges in the naive approach.
It is worth noting that the error generating procedures outlined in Example \ref{example::_gen_err_truncNormal} naturally ensure that $\bfeta$ satisfies the constrains in Section \ref{EE} for weakly exchangeable errors, thereby defining legitimate covariance matrices of errors on the parameter space.

In practice, to get the hybrid shorth estimate of $\eta_1$ from $\{\wh \zeta_{ij}\}$, we apply cross validation to tune the parameter $c$ \citep{Pensia2019entangled} as introduced in Section \ref{esteta}.
We set the possible range of $c$ to span from $2/\log(n^2-n)$ to $\sum_{i\neq j}^n\mathbb{I}\big[\wh\zeta_{ij} > - \max(\{\wh\zeta_{ij}\})\big]/\log(n^2-n)$ and denote the set of tunning parameters by $\mathcal{S}$. 
For each $c^* \in \mathcal{S}$, we randomly divide $\{\wh \zeta_{ij}\}$ into $10$ folds of approximately equal size. 
After selecting a validation set, we apply $k$-shorth method on the remaining $9$ folds. 
The mean squared error, $\text{MSE}_\ell$, $\ell = 1,2,\ldots,10$, is computed using the observations in the held-out fold and the $k$-shorth estimate. 
The positive $k$-shorth estimator $\eta_{1, +}$ is calculated by setting $c = \operatorname*{arg\,min}_{c^*} \{c^* \in \mathcal{S}: \frac{1}{10}\sum^{10}_{\ell=1}\text{MSE}_\ell\}$.
In practice, we enforce the positive semi-definiteness of $\wh \Omega_e$ by applying an eigenvalue correction to $\wh \eta_{1,\mathrm{hybrid}}$.
Specifically, we adjust the smallest eigenvalue of $\wh \bOmega_e$ to ensure it is nonnegative.
Our numerical experiments indicate that such a minor perturbation has a negligible impact on the computational accuracy of the final results.

\subsection{Comparison results of coverage probabilities}
\label{42}

In this section, we present simulation results for inferring regression coefficients applying settings in Example \ref{example::_gen_err_truncNormal}.
Additional studies with $\{e_{ij}\}$ from Gamma distribution and experiments with different covariate configurations are given in the Supplementary Material.

\begin{figure}[h]
    \centering
    \begin{tikzpicture}
        \node at (0,0) {\includegraphics[width=\linewidth]{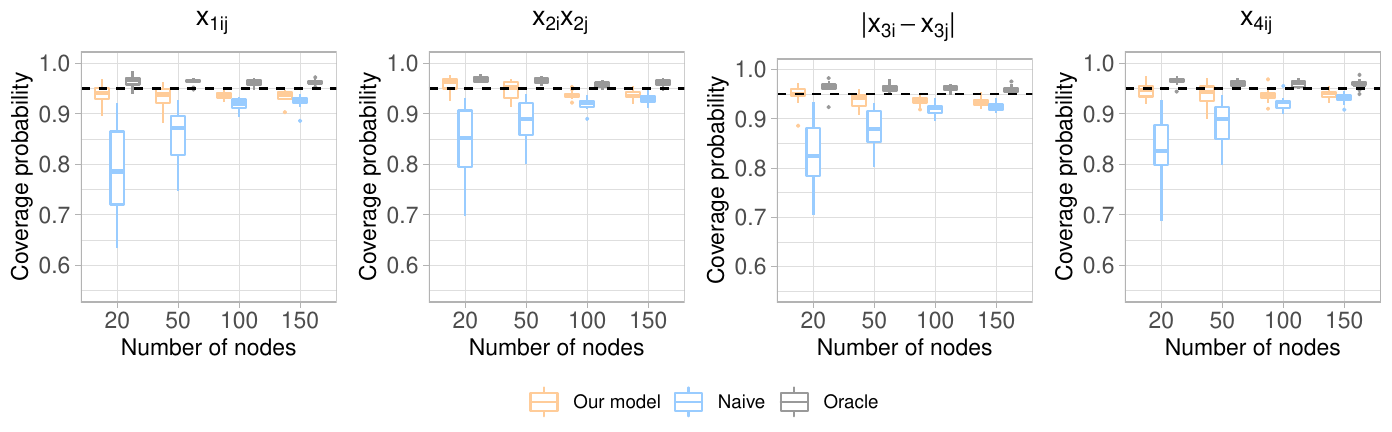}};
        \node at (0,-4.4) {\includegraphics[width=\linewidth]{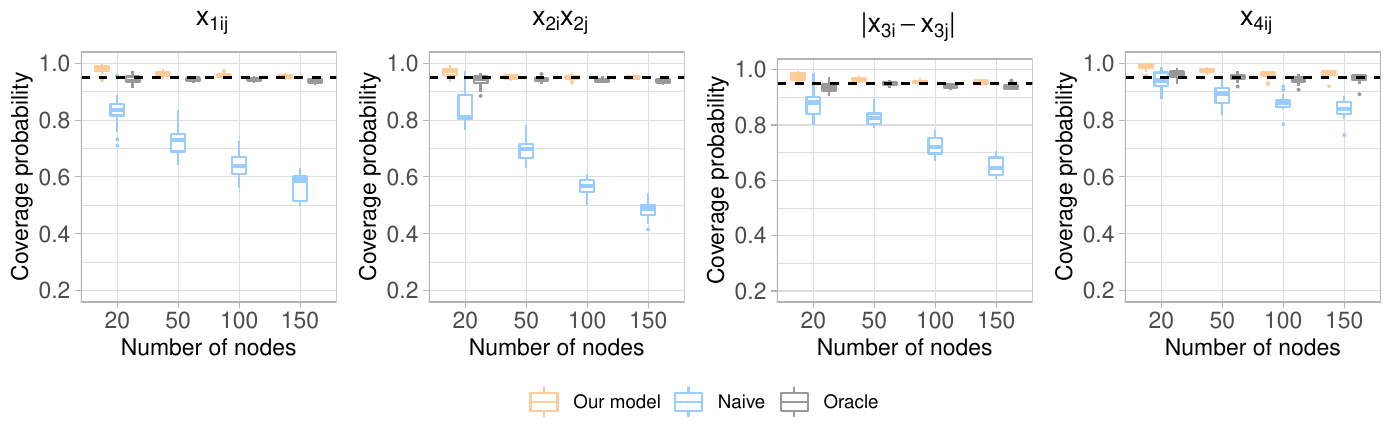}};  
    \end{tikzpicture}
    \caption {The empirical coverage probability of 95\% confidence interval of three competing methods under setting \eqref{gen1} (first row) and setting \eqref{gen2} (second row). } \label{figD2trN} 
\end{figure}

As shown in Figure \ref{figD2trN}, when error terms are independent and identically distributed, our method performs as good as the oracle results. 
The coverage probability of the naive method, however, is further from the nominal 95\% level, and its variability across different realizations is larger than that of our method, especially when the number of nodes is close to or less than 50. 
For dependent error terms generated from setting \eqref{gen2}, our method performs extremely well as it recovers the dependence structure in the relational data. 
Specifically, our proposed inference procedure produces confidence intervals with coverage probability close to the nominal 95\% level under all configurations. 
Its performance becomes better and closer to the oracle results as the size of relational data grows, whereas the coverage probability of naive method is far below the nominal level and becomes worse as the number of nodes increases.

In conclusion, our method outperforms the naive method across all settings, especially for weakly exchangeable dependent errors. 
Specifically, the empirical coverage probability of our method is approaching the nominal level and closely approximates the oracle benchmark as the number of nodes increases. 
Moreover, our method demonstrates robustness in terms of empirical coverage probability under different error generating procedure (heavy-tailed errors from Gamma distribution as well as light-tailed errors from truncated Normal distribution), and different configurations under model \eqref{simuModel}. 
Comprehensive simulation results further supporting these findings are relegated to the Supplementary Material.

\section{Food sharing network analysis}
\label{sec5_food}


In this section, we apply the proposed model to investigate the food sharing network introduced in Section \ref{sec1}. The data contains the number of transferred gifts over a yearlong period among 25 households of indigenous Mayangna and Miskito horticulturalists in Nicaragua, along with distance, relationship, and other covariates given in the Supplementary Material. The ``association index" \citep{cairns1987comparison} reflects the amount of time that households interact with one another, which characterizes the multi-faceted inter-household relationships. A complication which arises in the study is that not all households were present for the full duration of the yearlong study. For model interpretation, \cite{koster2014food} accounts for the variation in the proportion of the year for which both members of each dyad were simultaneously present in the community by entering the natural logarithm of this exposure as an offset variable. This modification allows us to model the expected number of gifts per year.
The social relations model (SRM) developed by \cite{kenny1984srm} is applied in \cite{koster2014food} to separate individual effects in the log mean structure from relationship effects in dyadic data. Their overall results indicate that food sharing networks largely correspond to kin-based networks of social interaction, suggesting that food sharing is embedded in broader social relationships between households. 

\subsection{Preliminary analysis}

Note that the marginal mean of $y_{ij}$ in our model does not depend on the network effects defined in Section \ref{secModel}. The reciprocity effect, same sender effect, same receiver effect, and sender-receiver effect defined in our model are characterized by $\bfeta$, which are different from the random effects defined in SRM. The analysis in \cite{koster2014food} assumes random effects are normally distributed, but the authors find a noteworthy outlier (the number of gifts given between Household 1 and Household 25) in the relationship-level random effects and therefore include a dummy variable to represent this outlier as a fixed effect. Our model, however, benefits from the flexibility where we do not introduce strict distributional assumption on error terms except for weak exchangeability. Therefore, we exclude the artificial relationship covariate between Household 1 and Household 25 as introduced in \cite{koster2014food}. 

We choose the exponential link function and put all variables in model \eqref{model1} for preliminary analysis. it is worth noting that household dyads which spend considerable time together typically have close kinship ties \citep{hames1987garden, alvard2009kinship, koster2014food}. This also agrees with the correlation of estimated coefficients given in the Supplementary Material, where we find strong correlation between the effect of association index and mother-offspring ties. Conceptually, association index is a proximity measure of close kin ties. To deal with the collinearity problem, we consider the model where mother-offspring, father-offspring, full sibling, or other close kin ties are omitted, whereas the association index ($\text{Association}_{ij}$) is reserved. We further consider a dummy variable $\text{Relatedness5}_{ij} = 1-\text{Relatedness1}_{ij}-\text{Relatedness2}_{ij}-\text{Relatedness3}_{ij}$, to denote weak ties as discussed in \cite{koster2014food}, given the definition of those attributes in the Supplementary Material. Note that fishing is a common strategy for virtually all households \citep{koster2014food}, and as mentioned in \cite{defrance2009zooarchaeology}, meat circulated as a source of wealth and people generated wealth from the products that animals produced. These findings suggest a potential colinearity between meat harvesting and the wealth of a household. Together with the observations according to results guven in the Supplementary Material, we omit ``Wealth" in the nodal covariates since there is notable correlation of its estimated coefficients with that of others (``Fish" and ``Pig"). Lastly, we omit ``Pastors" variable in our model since there is only 2 households with pastors among the 25. The structured sparsity introduced by it would make the inference procedure unstable.

\subsection{Model setting and estimation results}   

\begin{figure}[ht]
    \centering
    \includegraphics[width=\linewidth]{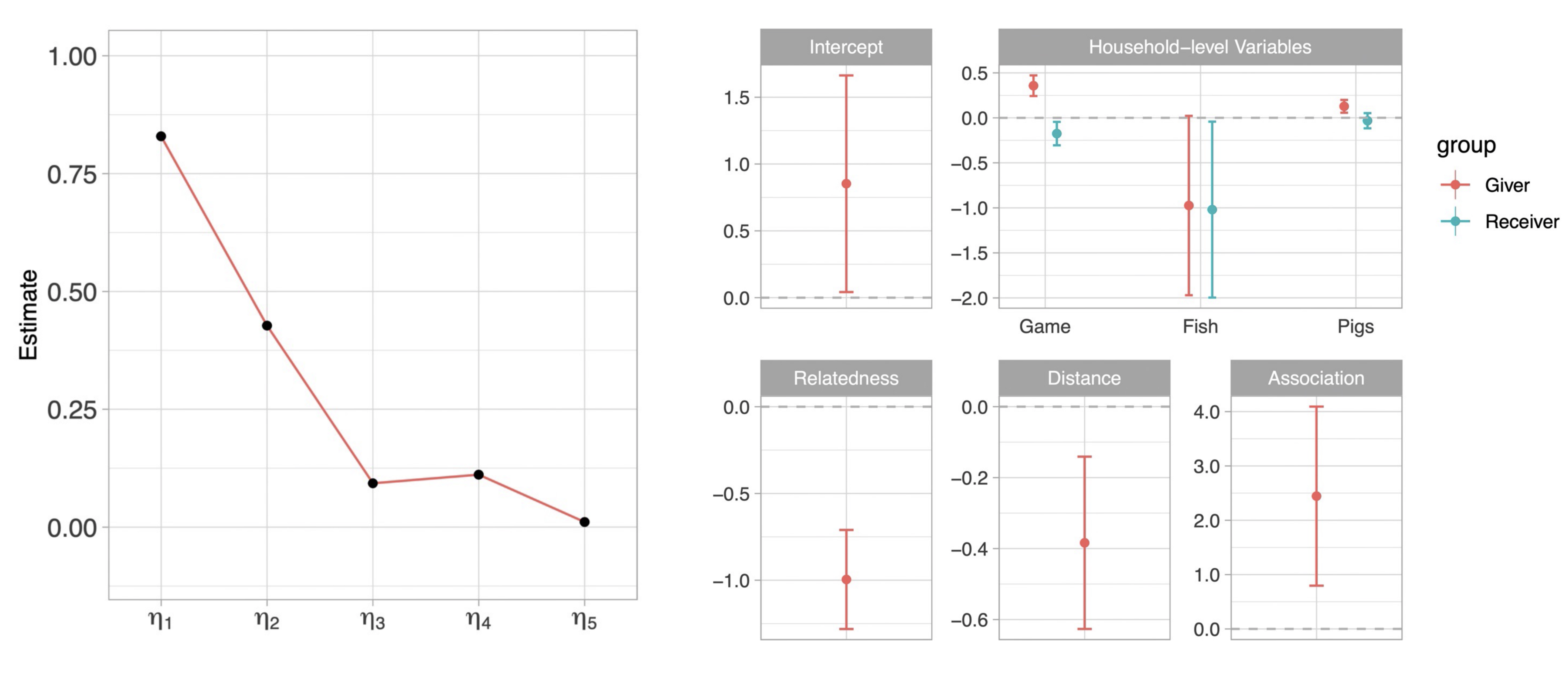} 
    \caption {Left: Point estimates of $\bfeta$. Right: Estimated regression coefficients and 95\% confidence intervals.} \label{food_reduced_CI} 
\end{figure}

We choose the exponential link function and put all variables in model \eqref{model1} for preliminary analysis. 
Pre-processing of the full sample is detailed in the Supplementary Material.
Our final model takes the following form: 
\begin{align*} 
    y_{ij} \sim ~  & \text{Poisson}\big[\exp\big\{ \beta_0 + \beta_1 \text{Game}_{i} + \beta_2 \text{Fish}_{i} + \beta_3 \text{Pigs}_{i} + \beta_4 \text{Game}_{j} + \beta_5 \text{Fish}_{j}\\ & + \beta_6 \text{Pigs}_{j} + \beta_7 \text{Relatedness5}_{ij} + \beta_8 \text{Distance}_{ij} + \beta_9 \text{Association}_{ij} \big\} e_{ij} \big],
\end{align*}
where $\{e_{ij}\}$ are weakly exchangeable with unit mean. In summary, after accounting for network effects, our model detects a statistically significant giver-game, giver-pigs, receiver-game, receiver-fish, weak kinship, distance, and association effects, but finds no evidence of effects for giver-game or receiver-pigs. It further suggests strong reciprocity effect and notable same sender/receiver effect in the relational data. 
Figure \ref{food_reduced_CI} shows the estimation results, where $\wh\bfeta = (0.829, 0.427, 0.093, 0.111, 0.011)$. 

Since the marginal mean of $y_{ij}$ in our model is different from that of SRM applied in \cite{koster2014food}, making it difficult to compare the regression coefficients of our model and theirs. Nonetheless, we can compare the significance of estimated coefficients of our model to the existing results. We can also verify whether the sign of $\eta_2$ through $\eta_5$ agrees with the result obtained in \cite{koster2014food} since the network effects can be presented by variance/covariance parameters in SRM model. Applying our definition of $\bfeta$ to SRM gives 
$\eta_2 = 0.643$; $\eta_3 = 0.756$; $\eta_4 = 0.426$; and $\eta_5 = 0.041$, where the variance/covariance parameters are defined in \cite{koster2014food}. Though not directly comparable, the positiveness of these four network effects defined in our model agree with the results in \cite{koster2014food}. 

{\it 6.3.1. Significant effects of distance and association index between households, together with giving and receiving behaviors.} As a result, our model finds 8 significant coefficients among the 10. The intercept is estimated to be 0.853, which is significantly different from 0, indicating the willingness of gift giving between the Households. Households who harvest more game ($\wh\beta_1=0.357$) and own more pigs ($\wh\beta_3=0.128$) are predicted to give significantly more gifts than households who harvest less game and own less pigs. However, there is no significant association between the amount of fish ($\wh\beta_2=-0.974$) that households harvest and the number of gifts they tended to give to other households, which partly covariates to the relatively small proportion of fish that are sent as gifts. Additionally, the association between the amount of pigs ($\wh\beta_6=-0.033$) the households own and gift receiving is not significant, having adjusted for the other factors in the model. 
Households located farther apart are predicted to exchange less gift ($\wh\beta_8=-0.384$) than nearby households. Distance is entered as a log-transformed variable and so its coefficient has a partial elasticity interpretation: a 10\% increase in the distance between two households is associated with a 3.8\% decrease in the expected number of gifts exchanged between the two households. Regarding the association index, households who associated more frequently with one another were predicted to give more ($\wh\beta_9=2.444$). Those results are similar to what is observed in \cite{koster2014food}. Moreover, our model predicts less transfers between households with weaker ties ($\wh\beta_7=-0.996$), after omitting the dummy variables denoting mother-offspring, father-offspring, full sibling, or other close kin ties. Similar observation on kin ties and food sharing could be found in \cite{helms1971asang} and \cite{parsons1974between}.

{\it 6.3.2. Different findings than prior research concerning the significant impact of receiving behavior on daily harvest of meat and fish.}
As for the difference in results, our model suggests that households who harvest less game ($\wh\beta_4=-0.175$) and less fish ($\wh\beta_5=-1.020$) receive significantly more gifts than households who harvest more game and fish, while \cite{koster2014food} finds no significant association between the amount of game/fish and the number of gifts they tended to receive from other households. This could result from that we omit ``Wealth" in our analysis due to collinearity while \cite{koster2014food} include this variable in their model. 

{\it 6.3.3. Implication of strong reciprocity effect and notable same sender/receiver effect.} Now, we analyze the dependence structure in the food sharing network. As depicted in Figure \ref{food_reduced_CI}, the reciprocity effect dominates the same sender effect, same receiver effect, and sender-receiver effect. 
Specifically, it suggests the gift giving behavior is reciprocated. 
Though relatively smaller compared to $\eta_2$, the magnitude of $\eta_3$ and $\eta_4$ are larger than $\eta_5$, indicating notable dependencies between relations involving the same sender or receiver. 

\section{Conclusion and discussion}
\label{sec6}

In this paper, we proposed a flexible multiplicative model on count edges for relational data. The model can handle different count distributions and is able to capture the underline pairwise dependence structure between edges given the observed data. For the regression setting, we proved that the proposed estimator is asymptotically normal and the estimate of covariance parameters is consistent, which delivers valid inference under the weak exchangeability assumption. Our work makes important progress toward the inference problem for modeling count edges in relational data. We also demonstrated the proposed model on a food sharing network.

Note that the estimation procedure of coefficients is robust to model misspecification if data comes from a linear exponential family with the same mean structure as in our model. The consistency of the estimator is guaranteed by the limit theory for the statistical agnostic, as discussed in \cite{Gourieroux1984pseudoMLE}. Under the model assumption of weakly exchangeable errors in Section \ref{EE}, the asymptotic variance of $\wh \bbeta_n$ is dominated by $\eta_{3}, \eta_{4}$, and $\eta_{5}$, since in $\bOmega_e$, elements of $\bfeta$ occur with multiplicity $n^2 - n$ (for both $\eta_{1}$ and $\eta_{2}$), $n^3 - 3n^2 + 2n$ (for both $\eta_{3}$ and $\eta_{4}$), and $2(n^3 - 3n^2 + 2n)$ for $\eta_{5}$. 
The variance-covariance estimation for Theorem \ref{co1} naturally involves $\wh\eta_1$ and $\wh\eta_2$ which accounts for the asymptotically negligible bias mentioned in \cite{graham2020dyadic}, though under a different modeling framework. As mentioned in \cite{graham2020cid}, the non-zero covariance terms are crucial for understanding the sampling distribution of $\wh \bbeta_n$. 
It is not of our interest in this work when the same sender effect, same receiver effect, and sender-receiver effect do not exist, though it gives a faster convergence rate of $\wh \bbeta_n$. 
Under this scenario, the asymptotic normality of $\wh \bbeta_n$ does not necessarily hold except for i.i.d. errors, as less correlation does not imply less dependencies in the model. 
Related discussions are available in \cite{menzel2021bootstrap}. 

Our work can be extended in several directions. One is adapting the model to relational data where the pairwise edge dependencies can vary with the sample size, potentially affecting the convergence rate of model estimators. 
Another promising direction is to study the dependence structure within the relational data, offering insights for hypothesis testing to compare dependencies across different datasets.

\bigskip
\begin{center}
{\large\bf SUPPLEMENTARY MATERIAL}
\end{center}

The complete proofs, additional simulation and empirical illustration results, and further discussions on the parameter space of covariance of weakly exchangeable variables can be found in the Supplementary Material. Code for reproducing the experiments and real data analysis can be obtained from \url{https://github.com/WenqinDu/Count_Relational_Data_Modeling}.

\bibliographystyle{apalike}
\bibliography{mybibliography}

\end{document}